\documentstyle[aps]{revtex}
\input psfig.tex
\begin{document}

\title{Hopping-disorder-induced effects upon
the two-Magnon Raman Scattering spectrum in an 
Antiferromagnet} 
\author{Saurabh Basu \thanks {e-mail : snbasu@iitk.ac.in}} 
\address{ 
Department of Physics, Indian Institute of Technology, Kanpur
208016, India}

\maketitle

\nopagebreak

\begin{abstract}
Two-magnon Raman scattering intensity is obtained 
for two-dimensional spin-$1/2$ antiferromagnets in presence of hopping 
disorder.  A consistent mode assignment scheme is prescribed which
essentially establishes a correspondence between 
the hopping-disordered and the pure system. 
It is seen that a minute amount of
disorder in hopping leads to a good agreement of the Raman lineshape
experimentally obtained for the copper-oxide insulators such as
${\rm {La_{2}CuO_{4}}}$. Also it is observed that a considerable asymmetry with
respect to the two-magnon peak appears and the spectral intensity
persisting much beyond $4J$ (where the joint magnon density of states
peaks). This has been argued earlier by us [S. Basu and A. Singh,
Phys. Rev. B, {\bf {53}}, 6406 (1996)] to be arising
due to highly asymmetric magnon-energy renormalization because of a
cooperative effect arising from local correlation in hopping disorder.

\ \\
\end{abstract}
\bibliographystyle{unsrt}



\section{Introduction}

Recently it was put forward by Nori et al. 
\cite{NORI} that the exchange disorder
caused by zero-point lattice vibration can account for various
experimentally observed features of the two-magnon Raman intensity
lineshape. In their approach the exchange disorder is looked upon as an
interaction between the lattice excitation (i.e. phonons) and the spin
excitations (i.e. magnons) which induces changes $\delta J_{ij}$, in the
exchange integral $J$. An estimate of the mean square deviation $\left \langle
(\frac {\delta J_{ij}}{J})^{2}\right \rangle$ is made and to make a comparison
with the experimental data the corresponding value is considered to be
as high as $0.5$ (in units of $J$). It has been suspected \cite {SCH} that such a large 
variation in $J$  appears physically unreasonable and questioned that
whether more moderate values of $\left \langle \frac {\delta J_{ij}}{J}
\right \rangle$ are sufficient to achieve agreement with experiments.
Essentially what one obtains is that the 
coupling of phonons to the spin excitations leads to the broadening
of a two-magnon peak, the asymmetry of the lineshape about its maximum
towards the higher energy regime and
appearance of scattering intensity in forbidden geometries like
$A_{1g}$, $B_{2g}$ etc. \\

Earlier, we have investigated, \cite{BS1} the effect of lattice fluctuation
induced hopping disorder in the context of Mott-Hubbard
AF where the hopping term $t_{ij}$ include random terms. 
The Hubbard Hamiltonian in this case can be written as,

\begin{equation}
\rm  {H}  =  - \sum_{\langle ij\rangle,\sigma}(t+\delta t_{ij})
(\hat {a^{\dagger}_{i\sigma}}\hat {a}_{j,\sigma} +{\rm {h.c.}}) + 
U\sum_{i}\hat{n}_{i\uparrow}\hat{n}_{i\downarrow} 
\end{equation}

\begin{equation}
P \left (\frac{\delta t_{ij}}{t}\right ) = \frac{1}{\sqrt{2\pi\sigma}}
exp\left [\frac{-(\delta t_{ij}/t)^{2}}{2\sigma}\right ]
\end{equation}

\noindent where the random terms $\delta t_{ij}$'s for each bond are chosen
independently from a gaussian distribution and the distribution width
$\sqrt{\sigma}$ measures the strength of disorder.
There we have examined both perturbatively and using the exact eigenstate
method - effects of hopping disorder on magnon energies, their wavefunctions
and on the density of states (DOS). In both  schemes, the spin-wave energies
are obtained at the Random Phase Approximation (RPA) level. We have obtained
a strong renormalization in energy of the high energy magnon modes, arising
from locally correlated hopping, which results in appreciable one-magnon
DOS well beyond the maximum spin-wave energy $2J$ for the pure system. This
result has got important implications in explaining the experimentally
observed Raman spectrum.\\

To obtain the Raman scattering intensity we need to consider
the imaginary part of the full two-magnon propagator which in turn
requires the knowledge of spin-wave energies. For a finite size system
these energies are obtained numerically at the RPA level
in presence of disorder using the exact eigenstate method discussed
in detail elsewhere \cite{BS3}. Following a 
prescription of the mode assignments discussed in section III, it is
possible to obtain the symmetry factor required for the $B_{1g}$ scattering
geometry (in Appendix A, we discuss the calculation of the
symmetry factor in the $B_{1g}$ geometry corresponding to the 
pure case). It is observed that even for a sufficiently small strength of
disorder, $\sigma = 0.01$, (much smaller than the 
values considered by Nori et al. \cite {NORI}) though the Raman
spectrum on the lower energy side is weakly affected,
there is a substantial change in the  
higher frequency regime and a good agreement is obtained with the
experimental data. This confirms our earlier statement \cite {BS1}
that if indeed the randomness in hopping due to zero-point lattice fluctuation
is playing a key-role in explaining the long standing puzzling feature
of the Raman lineshape regarding the asymmetry, then it 
essentially arises from the highly asymmetric magnon-energy 
renormalization.

\section{Configuration averaging of the two-magnon propagator}

The Fleury-Loudon
(FL) Hamiltonian which represents the interaction between the photon and the
spin pairs can be written as,

\begin{equation}
\rm {H_{R}} = A\sum_{{\bf {r}},{\hat {\delta}}} 
({\bf {E}}_{inc}.{{\hat {\delta}}})
.({\bf {E}}_{sc}.{{\hat {\delta}}}) {\bf S}({\bf {r}}).{\bf {S}}
({\bf {r}} + {{\hat {\delta}}})
\end{equation}

\noindent Where the ${\bf {E}}_{inc}$
and ${\bf {E}}_{sc}$ are the incident and
the scattered electric field vectors, 
${\hat {\delta}} $ is the unit vector
connecting nearest neighbour (NN) 
sites of opposite sublattices and $A$ is a constant.
The sum over ${\bf {r}}$ ensures that the spin
excitations have net zero momentum  
as the incoming photon has essentially zero momentum.
For calculation of the Raman intensity 
one needs to consider the correlation
function of the spin pair operator,

\begin{equation}
P_{\bf {\hat {\delta}}} = \sum_{\bf {r}} {\bf {S}}({\bf {r}}).{\bf {S}}
({\bf {r}}+{\hat {\delta}} )
\end{equation}

\noindent Now for the $B_{1g}$ symmetry we have,

\begin{displaymath}
{\bf {E}}_{inc} \sim \hat x - \hat y 
\end{displaymath} 

\begin{displaymath}
{\bf {E}}_{sc} \sim \hat x + \hat y. 
\end{displaymath} 

\noindent So that the Hamiltonian assumes the form,    

\begin{equation}
{\rm {H_{R}}} \sim  \sum_{{\bf {r}},p = \pm 1} 
{\bf {S (r)}}.\{{\bf {S}}({\bf {r}}+ p\hat x) - {\bf {S}}({\bf {r}} +p\hat y) \}
\end{equation}

Now let us consider the diagram representing  (Fig. 1)
the interaction between the magnons
in real space. The interaction between the magnons excited at
sites ${\bf {r_{1}}}$ and ${\bf {r_{1} + \mu}} $ can be written as,

\begin{equation}
{\hat {V}}_{\rm {int}} = -J\sum_{{\bf {r_{1},\mu}}}S^{-}({\bf {r_{1}}})
S^{+}({\bf {r}}_{1}) S^{-}({\bf {r}_{1}+\mu})
S^{+}({\bf {r}_{1}+\mu} )
\end{equation}

\noindent where $\mu $ connects to NN sites of the opposite sublattice.
The form of the interaction indicates that the  propagators
on either  sides of the interaction line are coupled  as $\mu $ can
connect both in the  $\hat {x}$ and  $\hat {y}$ direction.
But they can be decoupled if one rewrites
the interaction term as (see Fig. 1),
    
\begin{eqnarray}
{\hat {V}}_{\rm {int}} = -J \sum_{{\bf {r}}_{1}} 
S^{-}({\bf {r}}_{1})S^{+}({\bf {r}}_{1})
\biggl [ S^{-}_{O_{1}}({\bf {r}}_{1})S^{+}_{O_{1}}({\bf {r}}_{1}) 
& + & S^{-}_{O_{2}}({\bf {r}}_{1})S^{+}_{O_{2}}
({\bf {r}}_{1}) \\ \nonumber
+ S^{-}_{O_{3}}({\bf {r}}_{1})S^{+}_{O_{3}}({\bf {r}}_{1}) 
& + & S^{-}_{O_{4}}({\bf {r}}_{1})S^{+}_{O_{4}}
({\bf {r}}_{1})\biggr ]
\end{eqnarray}
    
\noindent  where the 4 operators are defined in the
following way,

\begin{displaymath}
S^{\pm}_{O_{1}}({\bf {r}}) = \frac{1}{2}\sum_{p} 
\{S^{\pm}({\bf {r}} + p\hat x) - S^{\pm}({\bf {r}} + p\hat y) \}
\end{displaymath}

\begin{displaymath}
S^{\pm}_{O_{2}}({\bf {r}}) = \frac{1}{2}\sum_{p} 
\{S^{\pm}({\bf {r}} + p\hat x) + S^{\pm}({\bf {r}} + p\hat y) \}
\end{displaymath}

\begin{displaymath}
S^{\pm}_{O_{3}}({\bf {r}}) = \frac{1}{2}\sum_{p} 
\{pS^{\pm}({\bf {r}} + p\hat x) - pS^{\pm}({\bf {r}} + p\hat y) \}
\end{displaymath}

\begin{displaymath}
S^{\pm}_{O_{4}}({\bf {r}}) = \frac{1}{2}\sum_{p} 
\{pS^{\pm}({\bf {r}} + p\hat x) + pS^{\pm}({\bf {r}} + p\hat y) \}
\end{displaymath}

\noindent where the $+$sign and the $-$sign in the superscript refers to 
raising and lowering operators respectively. 
The first two operators correspond to $B_{1g}$ and
$A_{1g}$ symmetry factors respectively.
Hence explicitly we write them as 
$ S^{\pm}_{B_{1g}}$ and $S^{\pm}_{A_{1g}}$  
for our convenience. In this new notation  the interaction term looks like,

\begin{eqnarray}
{\hat {V}}_{\rm {int}} = -J \sum_{{\bf {r}}_{1}} 
S^{-}({\bf {r}}_{1})S^{+}({\bf {r}}_{1}) \biggl
[ S^{-}_{B_{1g}}({\bf {r}}_{1})S^{+}_{B_{1g}}({\bf {r}}_{1}) 
& + & S^{-}_{\rm {A_{1g}}}({\bf {r}}_{1})S^{+}_{A_{1g}}
({\bf {r}}_{1}) \\ \nonumber
+ S^{-}_{O_{3}}({\bf {r}}_{1})S^{+}_{O_{3}}({\bf {r}}_{1}) 
& + & S^{-}_{O_{4}}({\bf {r}}_{1})S^{+}_{O_{4}}
({\bf {r}}_{1}) \biggr ]
\end{eqnarray}

\noindent Having written the interaction term we proceed to define the
spin-wave propagators in different scattering geometries. 
In Fig. 1 the spin raising and lowering
operators at ${\bf {r}}$ and ${\bf {r}} + \hat {\delta}$ are denoted by
$S^{+}({\bf {r}})$ and $S^{-}_{B_{1g}}({\bf {r}})$ and those at
at ${\bf {r'}}$ and ${\bf {r'}} + \hat {\delta'}$ are denoted by
$S^{-}({\bf {r'}})$ and $S^{+}_{B_{1g}}({\bf {r'}})$. This
immediately suggests that $\hat {\delta}$ and 
$\hat {\delta'}$ which connect NN pairs
are fixed in space and we can write the spin-wave propagators as,

\begin{equation}
\chi^{-+}({\bf {r,r'}}) = \langle \psi_{G}|S^{-}
({\bf {r}})S^{+}({\bf {r'}}) |\psi_{G} \rangle
\end{equation}

\noindent corresponding to the upper line and,

\begin{equation}
\chi^{+-}_{B_{1g}}({\bf {r ,r'}}) = \langle \psi_{G}|
S^{+}_{B_{1g}}({\bf {r}})S^{-}_{B_{1g}}({\bf {r'}})
|\psi_{G} \rangle
\end{equation}

\noindent corresponding to the lower one in Fig. 1. 
Now the time-ordered two-magnon propagator in the $B_{1g}$
geometry can be defined as,

\begin{equation}
G_{B_{1g}}(\omega ) = -i\int dt e^{i\omega (t-t')}
\langle \psi_{G}|T[\sum_{{\bf {r,r'}}}
S^{-}({\bf {r}},t)S^{+}_{B_{1g}}({\bf {r}},t)
S^{+}({\bf {r'}},t')S^{-}_{B_{1g}}
({\bf {r'}},t')]|\psi_{G} \rangle
\end{equation}

\noindent For calculating the Raman intensity first consider the
non-interacting limit of the two-magnon propagator, 

\begin{equation}
G^{0}_{B_{1g}}(\omega ) = i\int \frac{d\omega_{1}}
{2\pi } \sum_{{\bf {r,r'}}} \chi^{-+}({\bf {r,r'}},\omega_{1})
\chi^{+-}_{B_{1g}}({\bf {r  ,r'}},\omega - \omega_{1}) 
\end{equation}

The configuration averaged two-magnon Raman propagator is given by,

\begin{equation}
\overline {G^{0}_{B_{1g}}(\omega )} = i\int \frac{d\omega_{1}}
{2\pi } \sum_{{\bf {r,r'}}} \overline {\chi^{-+}({\bf {r,r'}},\omega_{1})
\chi^{+-}_{B_{1g}}({\bf {r,r'}},\omega - \omega_{1})}
\end{equation}

\noindent We can decouple the two propagators if we neglect 
the corrections arising due to exchange between the magnons. 
This allows us to write,

\begin{equation}
\overline {G^{0}_{B_{1g}}(\omega )} = i\int \frac{d\omega_{1}}
{2\pi } \sum_{{\bf {r,r'}}} \overline {\chi^{-+}({\bf {r,r'}},\omega_{1})} ~~
\overline {\chi^{+-}_{B_{1g}}({\bf {r,r'}},\omega - \omega_{1})}
\end{equation}

Having done the configuration averaging 
of $G^{0}(\omega )$ one can proceed to calculate the
two-magnon Raman intensity which is obtained from imaginary part of
$\overline {G(\omega )}$ where,

\begin{equation}
\overline {G(\omega )} =
\overline {\left [\frac{G_{0}(\omega )}{1 + J G_{0}(\omega )} \right ]}
\end{equation}

\noindent  We can neglect the vertex corrections since they appear 
only at the second order level and hence 
proportional to the fourth power of the 
disorder strength {\it {i.e.}} $\left (\frac{\delta t}{t}\right )
^{4}$ (in reference [5] this result is shown explicitly for 
the on-site (diagonal) disorder case and calculation of the vertex 
correction is discussed). So for weak
disorder, the correction will be insignificant. This enables to write,

\begin{equation}
\overline {G(\omega )} =
\left [\frac{\overline {G_{0}(\omega )}}{1 + J
\overline {G_{0}(\omega )}} \right ]
\end{equation}

Now it is shown in Appendix A 
that the $B_{1g}$ symmetry factor viz.  $(\cos q_{x} - \cos q_{y})^{2}$
corresponding to the pure case is obtained by Fourier transforming
the non-interacting two-magnon propagator 
$G^{0}_{B_{1g}}$ (Eq. 12).  This symmetry factor is important
for calculating the two-magnon Raman scattering intensity. 
It can also be noted that due to the form of the interaction i.e.
presence of both $B_{1g}$ and $A_{1g}$ terms in 
${\hat {V}}_{int}$, off-diagonal
combinations of different scattering geometries
in $G^{0}(\omega )$ are possible.
In that case the symmetry factor will contain contributions from different
geometries. We can see  it clearly if one considers,

\begin{math}
\langle S^{+}_{{\bf {r}}}(B_{1g}) S^{-}_{{\bf {r'}}}
(A_{1g}) \rangle
\end{math}
\noindent which when Fourier transformed yields,

\begin{equation}
\sum_{q} (\cos q_{x} -\cos q_{y})(\cos q_{x} +
\cos q_{y})\chi^{+-}(q,\omega ) \nonumber
\end{equation}

\noindent But this symmetry term can be shown to be identically equal to
zero for all frequencies because of the symmetry present in the magnetic
Brillouin Zone.  Hence the only terms which survive are the ones
that are diagonal in scattering geometries \cite{BSS}.\\

\section{Mode Assignments}

Here we want to establish a one-to-one correspondence between the pure
system and the one in presence of hopping disorder. 
For doing this we calculate
the magnon modes using the exact eigenstate method for a $10\times 10$
system in absence of disorder in the strong-coupling limit ($U/t = 200$) .
This procedure of obtaining the magnon modes numerically has been discussed
in reference [4]. These modes thus obtained are
assigned a pair of mode numbers $n_{x}$ and $n_{y}$ which are related the
wavevectors $q_{x}$ and $q_{y}$ via the relation 
$q_{x/y} = \frac{2\pi}{L}n_{x/y}$ in the translationally invariant
(pure) case.  The numbers $n_{x}$ and 
$n_{y}$ are chosen such that the corresponding
$q_{x}$ and $q_{y}$ are restricted to the upper-half (first two quadrants) of
the Brillouin Zone (BZ) where $q_{x}$ is allowed to have both positive
and negative values (ranging from $-\pi$ to $+\pi$) whereas $q_{y}$ 
is assigned only positive values (from $0$ to $\pi$). The other half of
the BZ does not need to be considered as the $q_{x}$ and $q_{y}$ values
(or equivalently the $n_{x}$ and $n_{y}$) lying within the region 
specified (see Fig. 2) are sufficient to label all the modes.
The Goldstone mode or the lowest energy mode is assigned with the mode
numbers $n_{x} = 0$, $n_{y} = 0$, the second collective mode with
$n_{x} = 1$, $n_{y} = 0$ (or equivalently $n_{x} = 0$, $n_{y} = 1$),
the third one with $n_{x} = 1$, $n_{y} = 1$ and so on.
Having done with the assigning scheme 
it is now possible to form the symmetry factor for the $B_{1g}$
scattering geometry which is  
$\phi_{n}^{2} = [\cos (\frac{2\pi}{L})n_{x} -\cos (\frac{2\pi}{L})n_{y}]^{2}$. 
This symmetry factor is necessary for calculating
the two-magnon Raman intensity. The consistency of the mode assigning scheme
is checked by re-evaluating the magnon energies using the mode 
numbers via the relation
$\omega_{n} = 2J\sqrt{1 - \gamma_{n}^{2}}$ where 
$\gamma_{n} = \frac{1}{2}[\cos (\frac{2\pi}{L})n_{x}
+\cos (\frac{2\pi}{L})n_{y}]$ and $J =\frac{4t^{2}}{U} = 0.02$ (i.e.
$ U/t = 200$). The energies
thus obtained show excellent agreement with the ones calculated from the
exact eigenstate method for the pure case.  
The same mode assignments are expected to be carried over for the intermediate
coupling regime. We show the comparison between the
energies calculated numerically from the exact eigenstate method
and the ones evaluated by using the  analytical expression
relevant to the intermediate-$U$ regime ($U/t = 10$) viz. 

\begin{equation}
\omega_{n} = 2J\left [(1 - \gamma^{2}_{n}) - \frac{t^2}{\Delta^2}
\{6+3 \cos\left (\frac{2\pi}{L}\right )n_{x}~ \cos\left (\frac{2\pi}{L}\right )
n_{y} -9\gamma^{2}_{n}\}\right ] ^{1/2}
\end{equation}

\noindent in Table I. We establish the correspondence between the pure and
the disordered system by assuming the same assigning 
scheme continues to work for the disordered system.
We show the mode energies for the disordered system ($\sigma = 0.1$) for 
a particular configuration in Table II. We also notice that the
degeneracy of the modes is lifted due to the presence of disorder and the
low energy modes are not significantly affected whereas there is a
considerable effect towards the higher energy side.
To substantiate
this result regarding the nature of the high energy modes, we have
examined the magnon wavefunction for the high-energy modes and find that
these modes are strongly localised in certain regions of the
lattice \cite{BS1}. In reference [3] it is also mentioned that the
localization of the high-energy magnon modes occurs in those regions of
the lattice where the locally-averaged hopping strength is significantly
higher than the bulk average, so that the local magnetization, which goes
as $\sim U/t$ is low. This is shown in Fig. 3.

\section{Calculation of the two-magnon Raman intensity}

As before, \cite{BS2} we can express the 
propagator in terms of it's non-interacting
components $G_{0}(\omega)$. The mode assignments allows us
to write $G_{0}(\omega)$ in  a $2\times 2$
matrix form with the mode energies and 
the symmetry factors are labeled by the 
mode numbers $n (=\sqrt{n^{2}_{x} + n^{2}_{y}})$. Hence we can 
sum up the perturbation series in powers of the
interaction between the magnons and which yields,

\begin{equation}
[G_{n}(\omega)] = \frac{[G_{0n}]}{{\bf 1} + J[G_{0n}]}
\end{equation}

\noindent As in reference [7] the above equation is written in terms 
of the matrix elements,
\begin{eqnarray*}
[G_{0n}(\omega)]_{AA}  & = & A_{n}, \\ \nonumber 
[G_{0n}(\omega)]_{BB}  & = & B_{n} ~~~~ {\rm { and}} \\ \nonumber
[G_{0n}(\omega)]_{AB}  & = & [G_{0n}(\omega)]_{BA} = ~~ C_{n} \\
\end{eqnarray*}

\begin{equation}
\overline {G_{n}(\omega )} = \overline {\frac{A_{n}+B_{n}+2C_{n} +
2J(A_{n}B_{n} -C^{2}_{n})}{(1+JA_{n})(1+JB_{n}) - J^{2}C^{2}_{n}}}
\end{equation}
\noindent where $A_{n}$, $B_{n}$ and $C_{n}$ are defined as,

\begin{eqnarray*}
&A_{n}& = \frac{1}{n} \sum_{n} m^{2}\phi^{2}_{n}\left 
(\frac{2J}{\omega_{n}}\right )^{2} \left [ a^{2}_{n} - \left 
(\frac{\omega_{n}}{2J}\right )^{2} -\left (\frac{a_{n}\omega }{2J}\right ) 
+ \frac{1}{2} \left 
(\frac{\omega }{2J}\right )^{2} \right ] \left [ \frac{1}{\omega -2\omega_{n}}
-\frac{1}{\omega +2\omega_{n}} \right ] \\ \nonumber
&B_{n}& = \frac{1}{n} \sum_{n} m^{2}\phi^{2}_{n}\left 
(\frac{2J}{\omega_{n}}\right )^{2}
\left [ a^{2}_{n} - \left (\frac{\omega_{n}}{2J}\right )^{2} 
+\left (\frac{a_{n}\omega }{2J} \right ) + \frac{1}{2}\left 
(\frac{\omega }{2J}\right )^{2} \right ] \left [ \frac{1}{\omega -2\omega_{n}}
-\frac{1}{\omega +2\omega_{n}} \right ] \\ \nonumber
&C_{n}& = \frac{1}{n} \sum_{n} m^{2}\phi^{2}_{n}\left 
(\frac{2J}{\omega_{n}}\right )^{2}
\left [b^{2}\gamma^{2}_{n} \right ] \left [ \frac{1}{\omega -2\omega_{n}}
-\frac{1}{\omega +2\omega_{n}} \right ]
\end{eqnarray*}
\noindent where,
\begin{quote}
$m = 1- 2\frac{t^{2}}{\Delta^{2}}$,~
$a_{n} = 1 - \frac{t^{2}}{\Delta^{2}} \left \{3 +\frac{3}{2}
\cos(\frac{2\pi }{L}n_{x})~
\cos(\frac{2\pi }{L}n_{y}) + \gamma_{n}^{2} \right \}$ ~~and \\
$b = 1- \frac{11}{2}\frac{t^{2}}{\Delta^{2}}$
\end{quote}

\noindent In Fig. 4 we have shown the Raman scattering spectra
in the $B_{1g}$ symmetry as a function of transferred photon
frequency $\omega$ (in units of $J$) for a system size $10\times 10$
in presence of disorder ($\sigma = 0.1$) and $U/t = 10$.
It is noticed that the linewidth to peak position ratio is even higher
than unity, whereas the experimentally value is closed to 0.38
\cite {RRP}. This
shows clearly that the strength of disorder $\sigma$ (or equivalently
the mean square deviation $\left \langle 
(\frac{\delta J_{ij}}{J})^{2}\right \rangle$)
does not need to be as large as 0.1 in order to get an agreement with
the experimentally observed data for the Raman spectrum. Fig. 5 shows
the same plot for more moderate values of $\sigma$ viz $\sigma = 0.01$ and
$0.03$. The spectral lineshape for the pure case ($\sigma = 0$) is
included for comparison.
This plot is reminiscent of the magnon DOS shown in Fig. 2 in 
reference [3].
There we have considered much larger strengths of disorder ($\sigma = 0.1$
and $0.3$).  It can be observed that the linewidth to peak
position ratio increases from 0.283 corresponding to the pure case
to 0.37 for $\sigma = 0.01$ and 0.410 for $\sigma = 0.03$ whereas
the position of the peak roughly stays the same in all the cases.
So it looks like $\sigma = 0.01$ roughly corresponds to the value
that is needed for parameterizing disorder strength  
in order to obtain agreement with the experimentally obtained Raman lineshape.
Also it is seen that in presence of disorder 
there is a pronounced asymmetry
in the higher frequency region of the spectrum and a considerable intensity is
observed all the way upto $6J$ and even more. This can 
be explained by arguing that the high
energy magnon modes are strongly affected due to a cooperative effect arising
from local correlations of hopping disorder, which resulted in an
appreciable one-magnon DOS beyond 
the maximum energy $2J$ for the pure system \cite {BS1}. 
So it seems that a much lesser 
value of the disorder strength can actually account for the
puzzling features of the Raman lineshape. Finally, in Fig. 6 we plot
the scattering spectrum in the strong coupling limit ($U/t = 100$)
for various disorder strengths  viz. $\sigma = 0.01, ~0.03$ and $0.1$ in order
to draw comparisons with the results obtained by Nori et al. \cite{NORI} since
their work deals with the strong coupling limit {\it {i.e.}}
$U/t \rightarrow \infty$. The linewidth to peak position
ratio in this case comes out to be $0.333, ~ 0.406$ and $1$ 
corresponding to $\sigma = 0.01, ~0.03 $ and $0.1$. Also there
is a pronounced asymmetry with respect to the two-magnon peak in the
higher frequency regime. To obtain an idea about the asymmetry
we measure the quantity $I(\omega = 4J)/I_{max}(\omega)$.
This has a value close to $0.4$ which agrees very well with the
corresponding quantity obtained from the 
experiments done on ${\rm {La_{2}CuO_{4}}}$ \cite{RRP}. Nori et al.
obtained agreement with experiments for $\sigma_{J} \sim 0.5J$ (they have
defined $\sigma_{J}$ in units of $J$) which is much more than what
we have considered. This can be explained by noting that their
Quantum Monte Carlo (QMC) calculations of the Raman spectrum for a 
$S = 1/2$ Heisenberg antiferromagnet in the $B_{1g}$ scattering
geometry reveals a much lesser linewidth to peak-position ratio ($\sim 0.176$)
since in their calculation the intrinsic magnon damping \cite{BS2} has not been
taken into account (remember we obtained $0.283$ for the 
corresponding quantity \cite{BS2} which is
much closer to the experimental value $0.38$ than that of Nori {\it {et al.}}).
This explains the reason why they need a much bigger width ($\sigma$) in the
distribution of the exchange constant $J$ so as to have an agreement 
with experiments. In fact, the evaluation of the different spectral moments
of the Raman lineshape due to Singh {\it {et al.}}\cite{RRP} shows the 
ratio of the second moment to that of the first one (which is like the
linewidth to peak-position ratio) as $0.23$  which tallies very well with the 
experimental value $0.28$.

In conclusion, we have studied the effects of hopping disorder on
the two-magnon Raman intensity using the magnon energies calculated
numerically using the exact eigenstate analysis.
The energies for the disordered case are assumed to have a one-to-one
correspondence with the ones in the pure case. This correspondence
is established by assigning mode numbers $n_{x}$ and $n_{y}$
which are related to the momentum labels via the relation
$q_{x/y} = (\frac{2\pi}{L})n_{x/y}$. The validity of the
assigning scheme is checked for the pure case by re-evaluating 
the mode energies using the mode numbers and hence comparing them 
with the ones obtained numerically. Once the consistency of
the assigning scheme is checked, the numbers can be used to 
determine the symmetry factors which are required for calculating
the Raman intensity in different scattering geometries.  
For the Raman lineshape in the $B_{1g}$
in presence of a small amount of disorder ($\sigma$ as small as
0.01) symmetry it is observed that the lineshape broadens 
and alongwith there appears a substantial asymmetry towards the higher 
frequency side while the low-energy modes are weakly affected.
It is also observed that the spectral intensity persists even beyond
$6J$ where the joint magnon DOS diverges at $4J$. We also obtained 
a good agreement for the linewidth and asymmetry
of the two-magnon profile observed experimentally from the Raman 
studies on cuprate insulators for
$\sigma = 0.01$.  This clearly reflects the role 
of zero-point fluctuations of the lattice to 
explain the anomalous features appearing in Raman spectrum for the
cuprate antiferromagnets via a highly 
asymmetric magnon energy renormalization
arising due to correlation of hopping disorder.


\section*{Appendix A}

Here we find an expression for the symmetry factor appropriate to the 
$B_{1g}$ scattering geometry.
The Fleury-Loudon Hamiltonian for the $B_{1g}$ scattering geometry is
written as (see Eq. 5), 

\begin{equation}
{\rm {H}} = A\sum_{{\bf {r}},p=\pm 1} {\bf {S}}({\bf {r}})
.\{{\bf {S}}({\bf {r}} +p\hat {x}) - {\bf {S}}({\bf {r}} +p\hat{y})\}
\end{equation}

\noindent The two-magnon propagator in the 
non-interacting limit for this case is,

\begin{equation}
G_{0}(\omega) = i\int\frac{d\omega_{1}}{2\pi}\sum_{\bf {r,r'}} 
\chi^{-+}({\bf {r,r'}},\Omega_{1})\chi^{+-}_{B_{1g}}
(\tilde {\bf {r}},\tilde {\bf {r'}},\omega - \omega_{1})
\end{equation}

\noindent where ${\bf {r}}$, ${\bf {r'}}$ denote 
lattice sites and $\tilde {\bf {r}} = {\bf {r}} +\hat {\delta} $,
$\tilde {\bf {r'}} = {\bf {r'}} +\hat {\delta'} $, 
where $\hat {\delta} /\hat {\delta'}$ connect to the sites
neighbouring to ${\bf {r}}$ and ${\bf {r'}}$ in all 
the four directions i.e. $\hat {\delta} /\hat {\delta'}$
can be $p\hat{x}$ and $p\hat{y}$ with $p = \pm 1$. Also the spin raising and
lowering operators forming the propagator is shown in the Fig. 7. \\

Clearly $\chi^{+-}_{B_{1g}}$ consists of 16 terms 
with $\hat {\delta} $ and $\hat {\delta'}$ 
having all possible combinations which are enumerated below (the energy
dependence is dropped temporarily).

\begin{eqnarray}
\chi^{+-}_{B_{1g}}(\tilde {\bf {r}},\tilde {\bf {r'}})  & = &
\chi^{+-}({\bf {r}}+\hat{x} , {\bf {r'}}+\hat{x}) +
\chi^{+-}({\bf {r}}+\hat{x} ,{\bf {r'}}-\hat{x}) \\ \nonumber
& - & \chi^{+-}({\bf {r}}+\hat{x} , {\bf {r'}}+\hat{y})- 
\chi^{+-}({\bf {r}}+\hat{x} , {\bf {r'}}-\hat{y})  \\ \nonumber
& + & \chi^{+-}({\bf {r}}-\hat{x} , {\bf {r'}}+\hat{x})+ 
\chi^{+-}({\bf {r}}-\hat{x} , {\bf {r'}}-\hat{x})  \\ \nonumber
& - & \chi^{+-}({\bf {r}}-\hat{x} , {\bf {r'}}+\hat{y})- 
\chi^{+-}({\bf {r}}-\hat{x} , {\bf {r'}}-\hat{y})  \\ \nonumber
& - & \chi^{+-}({\bf {r}}+\hat{y} , {\bf {r'}}+\hat{x})- 
\chi^{+-}({\bf {r}}+\hat{y} , {\bf {r'}}-\hat{x})  \\ \nonumber
& + & \chi^{+-}({\bf {r}}+\hat{y} , {\bf {r'}}+\hat{y})+ 
\chi^{+-}({\bf {r}}+\hat{y} , {\bf {r'}}-\hat{y})  \\ \nonumber
& - & \chi^{+-}({\bf {r}}-\hat{y} , {\bf {r'}}+\hat{x})- 
\chi^{+-}({\bf {r}}-\hat{y} , {\bf {r'}}-\hat{x})  \\ \nonumber
& + & \chi^{+-}({\bf {r}}-\hat{y} , {\bf {r'}}+\hat{y})+ 
\chi^{+-}({\bf {r}}-\hat{y} , {\bf {r'}}-\hat{y})  
\end{eqnarray}

\noindent To obtain the symmetry factor one has to Fourier transform all the
16 terms in $G_{0}(\omega)$. As an example, the calculations for the first
two terms are shown below, \\

\noindent (1) The first term is, \\

\begin{displaymath}
i\int\frac{d\omega_{1}}{2\pi}\sum_{\bf {r,r'}} 
\chi^{-+}({\bf {r,r'}})\chi^{+-}({\bf {r}}+\hat{x},{\bf {r'}}+\hat{x})
\end{displaymath}

\noindent Taking the Fourier Transform,

\begin{displaymath}
i\int\frac{d\omega_{1}}{2\pi}\sum_{\bf {r,r'}} \sum_{q,q'}
\chi^{-+}(q)e^{iq.({\bf {r-r'}})}\chi^{+-}(q')
e^{iq'.({\bf {r-r'}})}
\end{displaymath}

\noindent Which can be written as,

\begin{displaymath}
i\int\frac{d\omega_{1}}{2\pi} \sum_{q}
\chi^{-+}(q)\chi^{+-}(-q)
\end{displaymath}

\noindent (2) And the second term,

\begin{displaymath}
i\int\frac{d\omega_{1}}{2\pi}\sum_{\bf {r,r'}} 
\chi^{-+}({\bf {r,r'}})\chi^{+-}({\bf {r}}+\hat{x},{\bf {r'}}-\hat{x})
\end{displaymath}

\noindent  Fourier Transforming the above quantity yields,

\begin{displaymath}
i\int\frac{d\omega_{1}}{2\pi}\sum_{\bf {r,r'}} \sum_{q,q'}
\chi^{-+}(q)e^{iq.({\bf {r-r'}})}\chi^{+-}
(q')e^{iq'.({\bf {r-r'}}-2\hat{x})}
\end{displaymath}

\noindent Which is equivalent to,

\begin{displaymath}
i\int\frac{d\omega_{1}}{2\pi} \sum_{q}
\chi^{-+}(q)\chi^{+-}(-q)e^{-2iq\hat{x}}
\end{displaymath}

\noindent Likewise we can calculate the contributions from 
the other 14 terms and all of them combine to give,

\begin{equation}
G_{0}(\omega)=i\int\frac{d\omega_{1}}{2\pi}\sum_{q}
\chi^{-+}(q, \omega_{1})\chi^{+-}(-q,\omega-\omega_{1})
[(\cos q_{x} - \cos q_{y})^{2}]
\end{equation}

\noindent Hence the symmetry factor used in the $B_{1g}$ geometry is 
$(\cos q_{x} - \cos q_{y})^{2}$. By taking the appropriate signs (positive)
for the spin operators corresponding to the lower line in Fig. 7 for the 
$A_{1g}$ geometry, it is trivial to show that the symmetry factor is
$(\cos q_{x} + \cos q_{y})^{2}$. 

\newpage


\begin{figure}[p]
\centerline{\mbox{\psfig{file=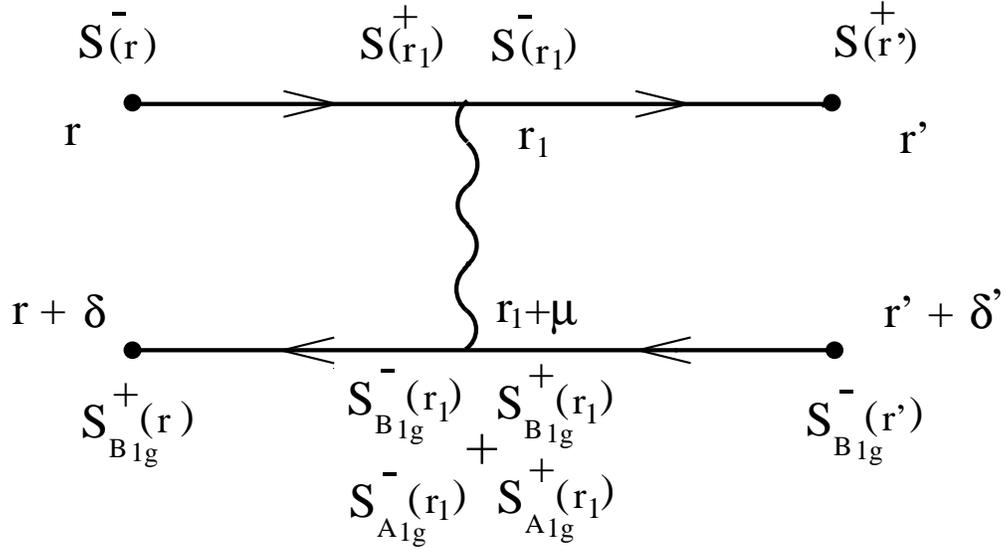,width=0.8\textwidth}}}
\caption{ Interaction between two magnons is shown
where the bold lines represent the
two magnons propagating from neighbouring sites ${\bf {r}}$ and 
${\bf {r}} +\delta $ to 
${\bf {r'}}$ and ${\bf {r'}} + \delta '$.The wavy lines denotes the interaction
between the magnons at sites ${\bf {r}}_{1}$ and ${\bf {r}}_{1} 
+ \mu $ where $\mu$
connects to the nearest neighbouring sites. The creation and destruction of
magnons at these points are shown by $S^{+}$ and $S^{-}$ 
where the subscripts $B_{1g}$ and $A_{1g}$ correspond to different scattering
geometries defined in the text.}
\end{figure}



\begin{figure*}[h!]
\centerline{\mbox{\psfig{file=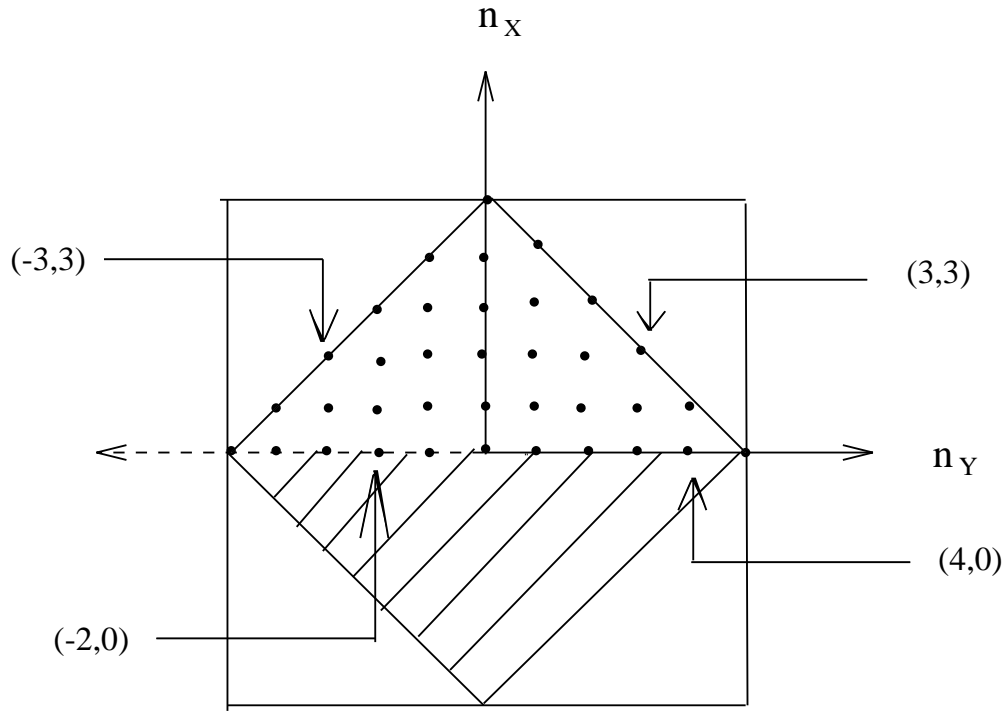,width=0.8\textwidth}}}
\caption{ The region in the Brillouin Zone (BZ) is shown where the 
dots denote the mode numbers $n_{x}$ and $n_{y}$ chosen so as to assign
labels to to the mode energies calculated. A few of the mode numbers are
indicated. The first two quadrants are required for assigning numbers to
all the modes.}
\end{figure*}

\begin{figure}[p]
\centerline{\mbox{\psfig{file=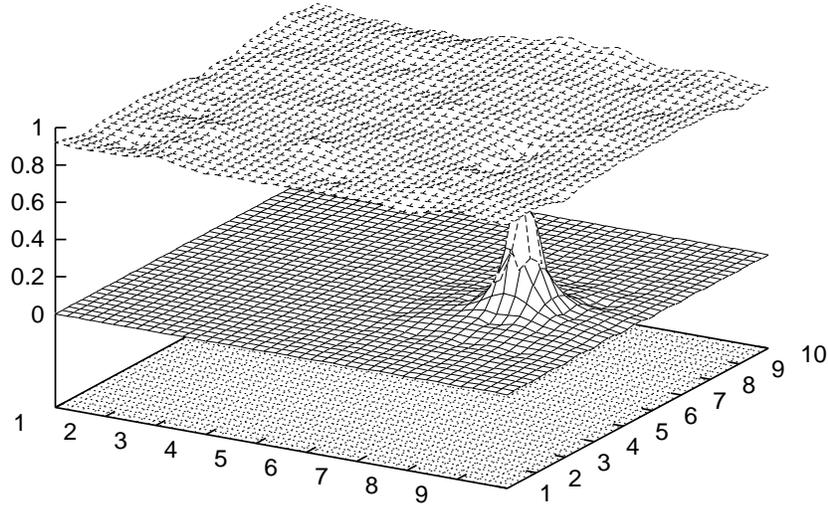,width=0.8\textwidth}}}
\caption{ The  amplitude for the highest energy magnon mode (below)
is shown for a $10 \times 10$ lattice for $U/t = 10$ in presence
of disorder $\sigma = 0.1$. Also we have shown the magnetization
profile (top) which shows a dip at places where the locally averaged
hopping strength is maximum.}
\end{figure}

\begin{figure}[p]
\centerline{\mbox{\psfig{file=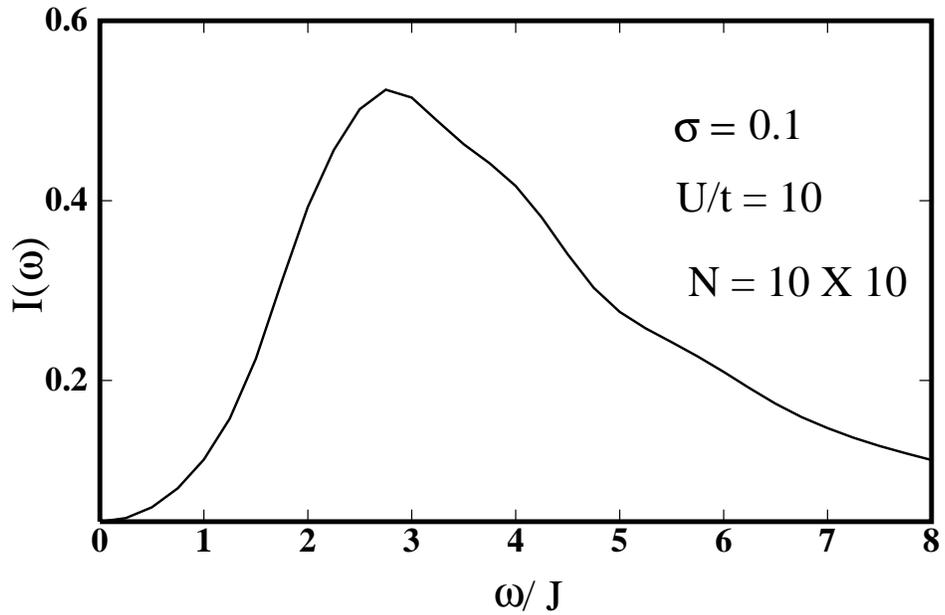,width=0.8\textwidth}}}
\caption{ The two-magnon scattering intensity is plotted as a 
function of transferred photon frequency $\omega$ (in units of $J$) for the
hopping disordered case with a gaussian disorder of width  
$\sigma = 0.1$.}
\end{figure}


\begin{figure}[p]
\centerline{\mbox{\psfig{file=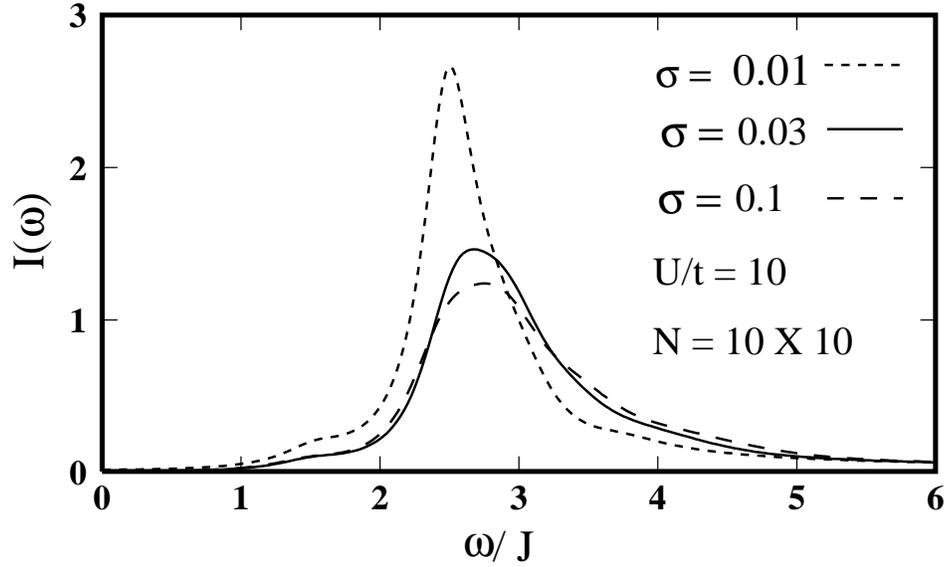,width=0.8\textwidth}}}
\caption{The two-magnon  Raman lineshape is shown 
for more moderate values of disorder, viz.
$\sigma = 0.01$ and $0.03$. The pure case ($\sigma = 0.0$) 
is included for comparison}
\end{figure}


\begin{figure}[p]
\centerline{\mbox{\psfig{file=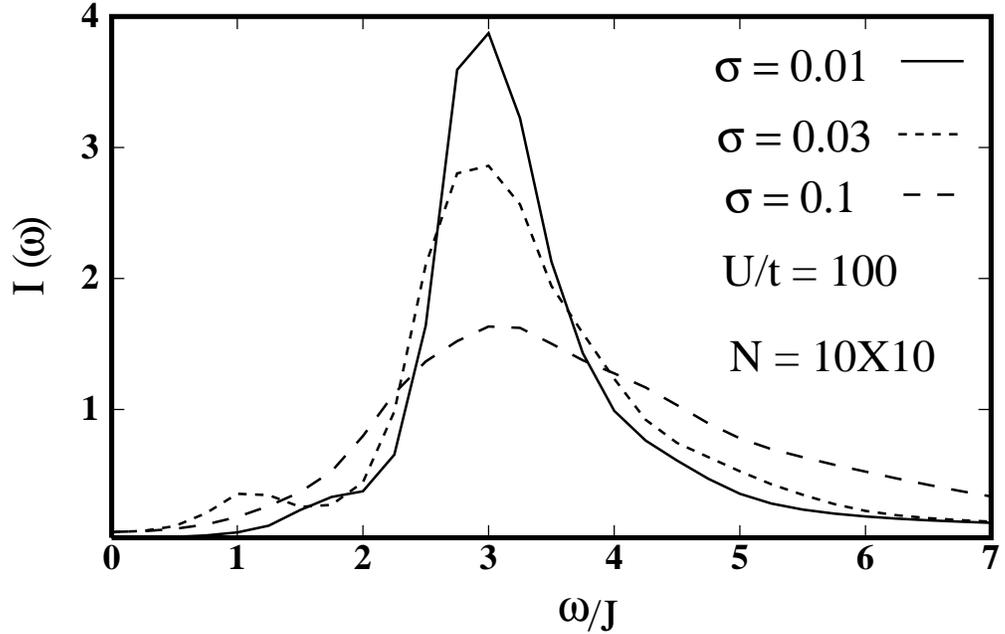,width=0.8\textwidth}}}
\caption{The two-magnon Raman lineshape is shown for the strong
coupling case  ($U/t = 100$) in order to compare with 
the results of Reference [1]. The values taken for disorder are
$\sigma = 0.01$, $0.03$ and $0.1$.  }
\end{figure}

\begin{figure}[p]
\centerline{\mbox{\psfig{file=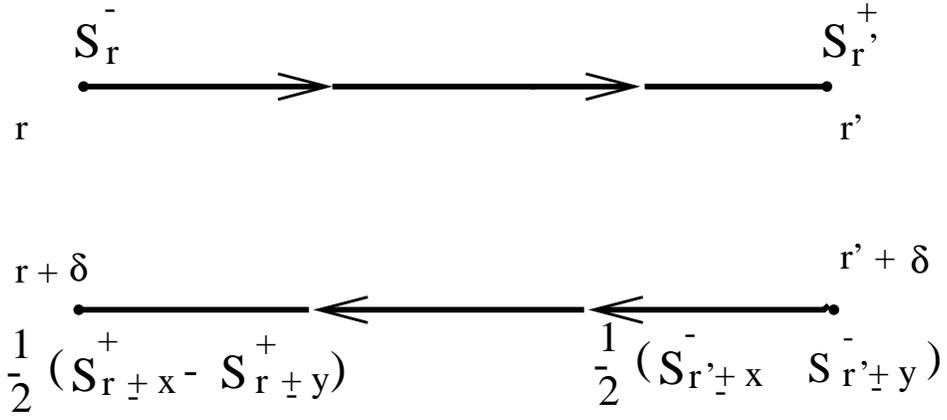,width=0.8\textwidth}}}
\caption{The two-magnon Raman propagator in the non-interacting limit 
corresponding to the $B_{1g}$ symmetry is shown.}  
\end{figure}


\newpage

\begin{table}
\caption{ The mode energies are obtained using the expression relevant
to the intermediate-$U$ regime ($U/t =10$ or $\Delta/t = 5$) for
a $10\times 10$ system where $n_{x/y} = (\frac{L}{2\pi})q_{x/y}$ are
the mode numbers specified in the leftmost column. Also we have shown the
energies calculated numerically using the exact eigenstate method
for comparison. The symmetry factors for both $B_{1g}$ and $A_{1g}$ 
scattering geometries are obtained using the mode numbers.}

\vspace{0.4in}

\begin{tabular}{ccccc} \hline \hline
Mode Nos. ($n_{x},n_{y}$)  & $\omega_{n}$ (numerical)  & 
$\omega_{n}$ (formula) & $\phi^{2}_{n}(B_{1g})$ & $\phi^{2}_{n}(A_{1g})$ \\ \hline
(-5,0)                      & 0.7548 & 0.7505 & 4.000 & 0.000  \\ \hline
(4,0) (0,4) (-4,0) (-5,1)   & 0.7438 & 0.7408 & 2.618 & 0.000  \\ \hline
(4,1) (1,4) (-4,1) (-1,4)   & 0.7396 & 0.7356 & 0.382 & 0.000  \\ \hline
(3,2) (2,3) (-3,2) (-2,3)   & 0.7150 & 0.6931 & 3.272 & 0.037  \\ \hline
(3,1) (1,3) (-3,1) (-1,3)   & 0.7030 & 0.6927 & 1.250 & 0.250  \\ 
(4,2) (2,4) (-4,2) (-2,4)   &        &        &       &        \\ \hline
(3,0) (0,3) (-3,0) (-5,2)   & 0.6882 & 0.6793 & 0.000 & 0.382  \\  \hline
(2,2) (3,3) (-2,2) (-3,3)   & 0.6765 & 0.6696 & 1.713 & 0.478  \\ \hline
(2,1) (1,2) (-2,1) (-1,2)   & 0.5945 & 0.5826 & 0.250 & 1.250  \\  
(4,3) (3,4) (-4,3) (-3,4)   &        &        &       &        \\ \hline
(2,0) (0,2) (-2,0) (-5,3)   & 0.5469 & 0.5458 & 0.477 & 1.714  \\ \hline
(1,1) (4,4) (-1,1) (-4,4)   & 0.4199 & 0.4102 & 0.000 & 2.618  \\ \hline
(1,0) (0,1) (-1,0) (-5,4)   & 0.3061 & 0.2987 & 0.036 & 3.273  \\ \hline
(0,0)                       & 0.0000 & 0.0000 & 0.000 & 4.000  \\ \hline \hline
\end{tabular}
\end{table}

\begin{table*}[h!]
\caption{Here the assignment of the mode numbers $n_{x}$ and $n_{y}$ is
shown for the mode energies calculated numerically for both pure and
disordered systems ($\sigma = 0.1$) of size $10\times 10$ and for $U/t = 10$.}

\vspace*{0.4in}

\begin{tabular}{ccccc} \hline \hline
Mode Nos. ($n_{x},n_{y}$)  & $\omega_{n}$ (pure case)  & 
$\omega_{n}$ (disordered case)  \\ \hline
(-5,0)                      & 0.7548 & 1.3006  \\ \hline
(4,0) (0,4) (-4,0) (-5,1)   & 0.7438 & 1.2222, 1.2024, 1.1960, 1.1346  \\ \hline
(4,1) (1,4) (-4,1) (-1,4)   & 0.7396 & 1.0578, 1.0178, 1.0137, 0.9955  \\ \hline
(3,2) (2,3) (-3,2) (-2,3)   & 0.7150 & 0.8557, 0.8414, 0.8291, 0.7934  \\ \hline
(3,1) (1,3) (-3,1) (-1,3)   & 0.7030 & 0.7854, 0.7762, 0.7698, 0.7641  \\ 
(4,2) (2,4) (-4,2) (-2,4)   &        & 0.7228, 0.7018, 0.6935, 0.6624  \\ \hline
(3,0) (0,3) (-3,0) (-5,2)   & 0.6882 & 0.6614, 0.6554, 0.6364, 0.6197  \\ \hline 
(2,2) (3,3) (-2,2) (-3,3)   & 0.6765 & 0.6020, 0.5773, 0.5668, 0.5411  \\ \hline
(2,1) (1,2) (-2,1) (-1,2)   & 0.5945 & 0.5251, 0.5167, 0.4926, 0.4784  \\  
			    &        & 0.4527, 0.4361, 0.4256, 0.4151  \\ \hline
(2,0) (0,2) (-2,0) (-5,3)   & 0.5469 & 0.3969, 0.3711, 0.3595, 0.3424  \\ \hline
(1,1) (4,4) (-1,1) (-4,4)   & 0.4199 & 0.3320, 0.3170, 0.2909, 0.2778  \\ \hline
(1,0) (0,1) (-1,0) (-5,4)   & 0.3061 & 0.2390, 0.2118, 0.1801, 0.1643  \\ \hline
(0,0)                       & 0.0000 & 0.0000  \\ \hline \hline
\end{tabular}
\end{table*}

\section*{Acknowledgment}

I am thankful to  Dr. Avinash Singh for his insightful comments
regarding  the manuscript. The work has been supported by a 
Research Grant SP/S2/M-25/95 from the
Department of Science and Technology (DST), India.

\end{document}